\documentstyle[12pt]{article}
\begin{document} 
\global\parskip 6pt
\newcommand{\be}{\begin{equation}}
\newcommand{\ee}{\end{equation}}
\newcommand{\bea}{\begin{eqnarray}}
\newcommand{\eea}{\end{eqnarray}}
\newcommand{\non}{\nonumber}

\begin{titlepage}
\hfill{hep-th/0106237}
\vspace*{1cm}
\begin{center}
{\Large\bf Brane World in a Topological Black Hole Bulk}\\
\vspace*{2cm}
Danny Birmingham\footnote{E-mail: dannyb@pop3.ucd.ie}
and Massimiliano Rinaldi\footnote{Email: massimiliano.rinaldi@ucd.ie}\\
\vspace*{.5cm}
{\em Department of Mathematical Physics\\
University College Dublin\\
Belfield, Dublin 4, Ireland}\\
\vspace{2cm}

\begin{abstract}
We consider a static brane in the background of a topological
black hole, in arbitrary dimensions. For hyperbolic horizons, we find
a solution only when the black hole mass assumes its minimum negative
value. In this case, the tension of the brane vanishes, and the
brane position coincides with the location of the horizon.
For an elliptic horizon, we show that the massless mode
of Randall-Sundrum is recovered in the limit
of large black hole mass.
\end{abstract}
\vspace{1cm}
July 2001
\end{center}
\end{titlepage}

\section{Introduction}
The possibility that our universe could be modelled as a brane
residing in a higher-dimensional anti-de Sitter spacetime has
received renewed impetus recently \cite{RS1,RS2}.
In this Randall-Sundrum scenario, the tension of the brane
is fixed in terms of the bulk cosmological constant, so that
the effective four-dimensional cosmological constant is zero.
A key point in the analysis is that although the extra dimension
is infinite in extent, the graviton fluctuations
include a normalizable zero mode which is localized near the brane.
In this way, one
can recover the usual gravitational physics of four
dimensions within a higher-dimensional context.

After these initial investigations, this scenario was generalized
in several ways. In particular, a static brane in
the background of a five-dimensional Schwarzschild-AdS black hole
was studied in
\cite{Gomez,Pasq}. In this case, the tension and position
of the brane are tuned in terms
of the black hole mass and cosmological constant.
The dynamics of a brane universe
in a black hole background has also been considered,
see \cite{Kraus,Tye,Csaki},
for example. 
Furthermore, the dynamics has been studied within the
context of the AdS/CFT correspondence.
One finds an interesting relation between the temperature and entropy
of the black hole and the Hubble parameter as the brane crosses the
horizon \cite{Savonije}. This analysis has also been extended
to the topological black hole case \cite{Cai,Youm}.

In this paper, we extend the discussion of a static brane
to the case of a topological black hole bulk, in arbitrary dimensions.
It is well known
that anti-de Sitter space affords the possibility of
black hole solutions, in which the topology of
the horizon is non-spherical
\cite{Beng}-\cite{Birmingham}.
The matching conditions at the brane fix both the brane tension
and the brane position in terms of the black hole mass and
cosmological constant \cite{Gomez,Pasq,Csaki}.
The key point for the case of hyperbolic black holes is the existence
of a negative mass spectrum.
We find a brane world solution only when the
black hole mass takes its minimum (negative) value.
In this case, the tension of the brane vanishes, and the brane position
coincides with the horizon. Unfortunately, the extent of the extra
dimension is then zero.
We also analyze the case of an elliptic horizon.
In particular, it is shown that the original Randall-Sundrum
graviton fluctuation equations are recovered in the limit of
large black hole mass.

\section{The Model}
We consider the following action \cite{RS1,RS2},
\bea
S = \int d^{d-1}x\;dz\; \sqrt{-G}(2 \mu^{d-2}R - \Lambda )
+ \int d^{d-1}x \sqrt{-g}(L - V),
\label{action}
\eea
where $\mu$ is the $d$-dimensional Planck mass, $L$  the Lagrangian
of the matter living on the brane, $V$ is the brane tension,
$\Lambda$ is the $d$-dimensional cosmological constant, and
$G_{MN}$ is the
$d$-dimensional metric with indices ($M,N=0,1,\ldots,(d-1)$).
The induced metric on the brane world
is denoted by $g_{\mu\nu}$, and is given by
$g_{\mu\nu}=G_{\mu\nu}(z = z_b)$, where $z_b$ is the position of the brane.
The coordinates of the bulk spacetime are denoted by
$x^{M} = (t, x^{i}, z)$, while the subset $x^{\mu} = (t, x^{i})$
are the coordinates of the brane world.
We shall consider the case with no matter on the brane, i.e.,
$L=0$. The corresponding equations of motion are given by
\bea
R_{MN}-\frac{1}{2}G_{MN}R = -\frac{1}{4\mu^{d-2}}
\left[\Lambda G_{MN} + V\frac{\sqrt{-g}}{\sqrt{-G}}
g_{\mu\nu}\delta^{\mu}_{M} \delta^{\nu}_{N} \delta(z-z_b)\right].
\label{einstein}
\eea

To begin, we consider the following ansatz for the bulk spacetime:
\bea
ds^{2} = -\left(k+\frac{r^2}{l^2}-\frac{2M}{r^{d-3}}\right)dt^2 +
\left(k+\frac{r^2}{l^2}-\frac{2M}{r^{d-3}}\right)^{-1}dr^2
+ r^2 \sigma_{ij}(x)dx^{i}dx^{j}.
\label{metric1}
\eea
This metric corresponds to a $d$-dimensional topological black
hole \cite{Birmingham}. In particular, the parameter
$k$ classifies the horizon topology, with $k=-1,0,+1$
corresponding to  hyperbolic, flat, and elliptic horizons,
respectively.
The five-dimensional case for $k=1$
was considered in \cite{Gomez,Pasq}.
The
parameter $M$ is identified with the mass of the
black hole, and $l$ is the radius of anti-de Sitter space.
In our conventions, the cosmological constant is given by
$\Lambda = -2(d-1)(d-2)\mu^{d-2}/l^{2}$.
Finally, $\sigma_{ij}$ is the metric
on the horizon of the black hole, which  we assume
to be an Einstein space with curvature $k$.

To proceed, it is convenient to change variables \cite{Gomez},
and consider
\bea
dz = \frac{1}{\sqrt{k+\frac{r^2}{l^2} - \frac{2M}{r^{d-3}}}} dr =
\frac{1}{A(r)}dr.
\label{def}
\eea
The metric then reads:
\bea
ds^{2} = -A(z)^{2} dt^{2} + B(z)^{2} \sigma_{ij}(x)dx^{i}dx^{j} + dz^{2},
\label{metric}
\eea
where $A(z)$ is defined by (\ref{def}), and $B(z)=r(z)$.
With this ansatz, the non-vanishing components of the Ricci
tensor are given by
\bea
R_{zz} &=& -\frac{A^{\prime\prime}}{A} -(d-2)\frac{B^{\prime\prime}}{B},
\non\\
R_{tt} &=& AA^{\prime\prime} + (d-2) \frac{AA^{\prime}B^{\prime}}{B},
\non\\
R_{ij} &=& R_{ij}(\sigma) - \sigma_{ij}\left(BB^{\prime\prime} +
\frac{BB^{\prime}A^{\prime}}{A} + (d-3)B^{\prime 2}\right).
\eea
Here, the prime denotes differentiation with respect to $z$,
and $x^{i}$ with $(i=1,...,(d-2))$ denote
the spatial coordinates of the horizon.
Since the horizon is an Einstein space, we have \cite{Birmingham}
\be
R_{ij}(\sigma) = (d-3)k\; \sigma_{ij}.
\ee
Hence, the Ricci scalar curvature is given by
\bea
R = (d-2)\left[ (d-3) \frac{k}{B^2} - 2\frac{A^{\prime}B^{\prime}}{AB}
- 2 \frac{B^{\prime\prime}}{B} - (d-3)\frac{B^{\prime 2}}{B^2}\right]
-2 \frac{A^{\prime\prime}}{A}.
\label{trace}
\eea
The $zz$, $tt$, and $ij$ equations of motion are given respectively by
\bea
0 &=& -k +
\left(\frac{2}{d-3}\right)\frac{A^{\prime}B^{\prime}B}{A}
+ B^{\prime 2} + \frac{\Lambda}{2\mu^{d-2}}
\frac{B^{2}}{(d-2)(d-3)},\label{zz}\\
0 &=& k - \frac{2}{(d-3)}B B^{\prime\prime} - B^{\prime 2}
- \frac{B^2}{2 \mu^{d-2}(d-2)(d-3)}(\Lambda
+ V\delta(z - z_{b})),\label{tt}\\
0 &=& \frac{(d-3)}{2}\left[- (d-4)k + (d-4)B^{\prime 2}
+ 2 B B^{\prime\prime} + 2\frac{A^{\prime}B^{\prime}B}{A}\right]
+ \frac{B^{2} A^{\prime\prime}}{A}\non\\
&+& \frac{B^2}{4\mu^{d-2}}(\Lambda + V\delta(z-z_b)).
\label{ij}
\eea
One sees immediately that the latter two equations
depend on the brane tension $V$.
It is also straightforward to check
that these equations hold in the bulk \cite{Birmingham},
when $V=0$.

\section{Matching Conditions}

In the original Randall-Sundrum model \cite{RS1},
the extra dimension is chosen to be periodic with an
orbifold structure, corresponding to a space
$S^1/ \mathbf{Z}_2$.
In our case, we follow this suggestion by cutting off the
extra coordinate at the horizon of the black hole ($z_{H}$) and at
the brane position ($z_{b}$), see \cite{Gomez,Csaki}. The
spacetime in between will then be referred to
as the bulk. Then, by a similar orbifold procedure, we
replace the cut part by a symmetric copy of the bulk.
This can be explicitly performed by the change of variables \cite{Gomez},
$\bar{z} = z_{b} - z$, with $z_{H} \leq z \leq z_b$, i.e.,
we consider only the spacetime between the horizon and the brane.
Next, we identify $\bar{z} \rightarrow -\bar{z}$.
The line element then becomes
\be
ds^{2} = -A(\bar{z})^{2}dt^{2} + B(\bar{z})^{2} \sigma_{ij}(x)dx^{i}dx^{j}
+ d{\bar{z}}^{2}.
\ee
Note that, in these new coordinates, $\bar{z} = 0$ corresponds to
the brane position, while $\bar{z}_{H} \equiv z_{b} -z_{H}$
is the position of the horizon. The metric
components are continuous at
$\bar{z} = 0$, but their first derivatives are not.
Therefore, any second derivative carries a singular
behaviour at the brane position.
Thus,
\be
A^{\prime\prime}(|\bar z|) = A^{\prime\prime}(\bar z)
+ 2A^{\prime}(\bar z)\delta(\bar z).
\ee
There is no matching condition at the location of the horizon,
as follows by definition from (\ref{def}).

Equations (\ref{tt}) and (\ref{ij}) then yield two conditions
on the brane tension and brane position.
This leads to a specification of $V$ and $r(0)$
in terms of the the black hole mass $M$, the cosmological constant $l$,
and the topological label $k$.
We find that
\be
V = 8(d-2)\mu^{d-2}\sqrt{\frac{k}{r(0)^2}
+\frac{1}{l^2}-\frac{2M}{r(0)^{d-1}}},
\ee
with the position of the brane given by
\be
r(0)^{d-3} = \frac{(d-1)M}{k}.
\label{position}
\ee
By combining these, the brane tension can be expressed in the
form
\be
V = 8(d-2)\mu^{d-2}\left[\frac{1}{l^2}
+ (d-3) \left(\frac{k}{d-1}\right)^{\frac{d-1}{d-3}}
\frac{1}{M^{\frac{2}{d-3}}}
\right]^{1/2}.
\label{tension}
\ee
According to the three different topologies,
we have the following scenario.
\begin{itemize}
\item For $k=-1$, the mass spectrum
includes black holes with negative mass, $M \geq M_{crit}$, where
\cite{Birmingham}
\be
M_{crit} = -\frac{l^{d-3}}{d-1}\left(\frac{d-3}{d-1}\right )^{(d-3)/2}.
\ee
The corresponding horizon location is denoted by $r_{crit}$, and given by
\be
r_{crit}=l\left (\frac{d-3}{d-1}\right )^{1/2}.
\ee
One can re-write (\ref{tension}) in the form
\be
V = 8(d-2)\frac{\mu^{d-2}}{l}\left[1
- \left(\frac{M_{crit}}{M}\right)^{\frac{2}{d-3}}\right]^{1/2}
\label{tension2}
\ee
Clearly, a real-valued brane position is possible by choosing
$M <0$ in (\ref{position}).
However, it then follows from (\ref{tension2})
that the only solution with real-valued brane tension
corresponds to $M = M_{crit}$.
Moreover, the brane tension then vanishes,
and the position of the brane coincides with the
location of the horizon $r(0) = r_{crit}$.
Unfortunately, the range of the extra coordinate is also zero,
and consequently this case becomes degenerate.

\item For $k=0$, the only solution is when $M=0$. In this case,
we recover the original Randall-Sundrum model \cite{RS2} for $d=5$.

\item For $k=1$, we have a range of solutions depending on the mass of the
black hole. The brane tension is the same
as that found in the original Randall-Sundrum model \cite{RS1,RS2},
plus a correction proportional to an inverse power of the mass.
Therefore, in the zero mass limit the
tension diverges, while in the large $M$ limit we recover the
Randall-Sundrum brane tension.
In this large $M$ limit, the brane runs off
to infinity faster than the horizon. For example, in five
dimensions \cite{Pasq}, one has
$r(0) \propto M^{1/2}$ and $r_H \propto M^{1/4}$.
Therefore, an observer living on the brane will find himself
in the asymptotic region of the black hole.
\end{itemize}

\section{Perturbations}
In this section, we analyze the graviton propagation for the
$k=1$ case \cite{Gomez}.
In particular, we show that the massless mode of Randall-Sundrum
is recovered in the large mass limit.
We first note that eqn. (\ref{einstein}) can written in the form
\be
R_{MN}=-\frac{(d-1)}{l^2}G_{MN}+T_{MN},
\label{einstein1}
\ee
where
\be
T_{MN}= \frac{V}{4\mu^{d-2}}\frac{\sqrt{-g}}{\sqrt{-G}}
\delta(z-z_{b})\left[\frac{(d-1)}{(d-2)}G_{MN} -  g_{\mu\nu}
\delta_{M}^{\mu}\delta_{N}^{\nu}\right].
\ee
If we introduce a small perturbation $h_{MN}(z,x^{\mu})$,
with $h_{z\mu} = h_{zz}=0$, then eqn. (\ref{einstein1}) reduces to
\be
R_{MN}^{(1)}=-\frac{(d-1)}{l^2}h_{MN} + T_{MN}^{(1)},
\label{pert}
\ee
with
\be
R_{MN}^{(1)}=\frac{1}{2}\left(-\nabla_{M}\nabla_{N}h
-\nabla^{2}h_{MN}+ \nabla^{A}\nabla_{M}h_{NA}
+ \nabla^{A}\nabla_{N}h_{MA}\right),
\label{h}
\ee
and
\be
T^{(1)}_{MN}=\frac{V}{4(d-2)\mu^{d-2}}\delta(z-z_{b})h_{MN}.
\ee
By commuting the covariant derivatives in (\ref{h}), and
imposing the Lorentz gauge $\nabla^{M}h_{MN}=0$, we find that
the perturbation equation (\ref{pert}) becomes
\be
\nabla_{M}\nabla_{N}h+\nabla^{2}h_{MN}- R_{AMNB}h^{AB}
-R_{ANMB}h^{AB} = 0.
\label{pert1}
\ee

In our case, the spacetime is warped asymmetric, with $A(z)$
different from $B(z)$.
If we assume that
the perturbation is traceless, $h^{M}_{\phantom{M}M} = 0$,
then the $zz$, $zt$, and
$zi$ equations are given respectively by
\bea
\left(\frac{B^{\prime 2}}{B^{2}} - \frac{B^{\prime\prime}}{B}\right)
h^{l}_{\phantom{l}l} +
\left(\frac{A^{\prime 2}}{A^{2}} - \frac{A^{\prime\prime}}{A}\right)
h^{t}_{\phantom{t}t} &=& 0,\non\\
\left(\frac{A^{\prime}}{A} - \frac{B^{\prime}}{B}\right)
\partial_{t}h_{tt} = 0,\;\;
\left(\frac{A^{\prime}}{A} - \frac{B^{\prime}}{B}\right)
\partial_{t}h_{ti} &=& 0.
\eea
Furthermore, the remaining $tt$, $ti$, and $ij$ equations
are of the form
\bea
0 &=& \left[ \partial_{z}^{2} + G^{tt}\partial_{t}^{2}
+ G^{ij}\nabla_{i}\nabla_{j} +
\left((d-2)\frac{B^{\prime}}{B}
- 3 \frac{A^{\prime}}{A}\right)\partial_{z} -
2 \frac{A^{\prime\prime}}{A} + 2 \frac{A^{\prime 2}}{A^{2}}
- 2(d-3)
\frac{A^{\prime}B^{\prime}}{AB}\right] h_{tt},\non\\
0 &=& \left[\partial_{z}^{2} + G^{tt}\partial_{t}^{2} + G^{jk}
\nabla_{j}\nabla_{k} +
\left((d-4)\frac{B^{\prime}}{B}
- \frac{A^{\prime}}{A}\right)\partial_{z} -
\frac{A^{\prime\prime}}{A} - \frac{B^{\prime\prime}}{B}
- (d-5)\frac{A^{\prime}B^{\prime}}{AB}\right. \non\\
&-& \left.(d-3)\frac{B^{\prime 2}}{B^{2}} \right]h_{ti},\non\\
0 &=& \left[\partial_{z}^{2} + G^{tt}\partial_{t}^{2} +
G^{kl}\nabla_{k}\nabla_{l}
+ \left(\frac{A^{\prime}}{A} + (d-6)\frac{B^{\prime}}{B}\right)
\partial_{z}
- 2 \frac{B^{\prime\prime}}{B} - 2 \frac{A^{\prime}B^{\prime}}{AB}
- 2 (d-5)\frac{B^{\prime 2}}{B^{2}}\right. \non\\
&-& \left. \frac{2}{B^{2}}\right]h_{ij}
- 2\left(\frac{B^{\prime 2}}{B^{2}} - \frac{A^{\prime}B^{\prime}}{AB}
- \frac{1}{B^{2}}\right)h^{l}_{\phantom{l}l}G_{ij}.
\eea
For the symmetric case, $A(z) = B(z)$, these equations agree with
\cite{Alonso}.

One can now examine these equations in the near-brane region,
by expanding the coefficients around the brane position.
In the limit of large black hole mass, the fluctuation
equations reduce to the single equation
\be
\left[\partial_{z}^{2} + G^{tt}\partial_{t}^{2}
+ G^{ij} \nabla_{i}\nabla_{j} - (d-5)\frac{1}{l}\partial_{z}
- 2(d-3)\frac{1}{l^2}
+ \frac{4}{l}\delta(z - z_{b})\right]h_{\mu\nu}=0.
\label{limit}
\ee

It is instructive to analyze the five-dimensional case in
more detail.
The gravitational equations on the $3$-brane have been studied in
\cite{Shiromizu}, see also \cite{Deruelle}. In particular,
the four-dimensional effective vacuum energy
(in 4-dimensional Planck units),
and the effective Planck mass are given by \cite{Csaki,Shiromizu}
\be
\Lambda_{4} = \frac{3k^2}{8M},\;\;
M^{2}_{PL} = \mu^{3}l\left(1 + \frac{l^2}{8M}\right)^{-\frac{1}{2}}.
\ee
Thus, we see that $\Lambda_{4}$ tends to zero for large mass, while
$M^{2}_{PL}$ tends to $\mu^{3}l$. The latter
corresponds to the Randall-Sundrum value \cite{RS2}
of the Planck mass, when the compactification radius is infinite.
In this sense,
one can regard the black hole mass parameter as being analogous
to the compactification radius of the Randall-Sundrum model.
Since $\Lambda_{4}$
vanishes in the large mass limit, we see that eqn. (\ref{limit})
does indeed coincide with the Randall-Sundrum
equation for graviton fluctuations.

In conclusion,  we have considered a static brane
in the background of a $d$-dimensional topological
black hole. One finds that both the tension and position
of the brane are fixed by the matching conditions in Einstein's equations.
For the case of $k=-1$, there is a solution only when the black hole
mass takes its lowest (negative) value. However,
since the brane is located at the horizon, the extra dimension does not
have extent.
For the $k=1$ case, it is in the limit of large black hole mass
that one recovers
the original Randall-Sundrum model, with a massless localized graviton.
Aspects of brane worlds in black hole backgrounds of higher derivative 
gravity have been studied in \cite{Odin1,Odin2}. 
Recently, various brane solutions in charged dilatonic backgrounds
have been discussed, and the $k=0$ and $k=-1$ solutions have particularly
interesting properties \cite{Grojean}.

\noindent {\bf Acknowledgements}\\
The research of M.R. is supported by
Enterprise Ireland grant BR/1999/031, and P.E. 2000/2001 from Bologna
University.
M.R. is grateful to P.J. Silva for valuable correspondence,
and to R. Balbinot for discussions.

\end{document}